\documentclass[prb,aps,floatfix,onecolumn,superscriptaddress,showpacs,amsmath,amssymb]{revtex4-2}
\usepackage{graphicx}
\usepackage{epstopdf}
\usepackage{epsfig}
\usepackage{epsf}
\usepackage{url}
\usepackage[USenglish]{babel}
\usepackage{hyperref}
\allowdisplaybreaks
\usepackage{verbatim}
\usepackage{natbib}
\setcitestyle{numbers}
\usepackage{amssymb,mathrsfs}
\usepackage{amsmath, nccmath}
\usepackage[export]{adjustbox}
\usepackage{dcolumn}
\usepackage{bm}
\usepackage{bbm}
\usepackage{lipsum}
\usepackage{multirow}
\usepackage{xcolor}
\usepackage[section]{placeins}

\newcommand{\tr}{\operatorname{Tr}}
\begin{document}
\title{Multidimensional coherent spectroscopy of correlated lattice systems}

\author{Jiyu Chen}
\affiliation{Department of Physics, University of Fribourg, 1700 Fribourg, Switzerland}
\author{Philipp Werner}
\affiliation{Department of Physics, University of Fribourg, 1700 Fribourg, Switzerland}
	
\begin{abstract}		
Multidimensional coherent spectroscopy (MDCS) has been established in quantum chemistry as a powerful tool for studying the nonlinear response and nonequilibrium dynamics of molecular systems. More recently, the technique has also been applied to correlated electron materials, where the
interplay of localized and itinerant states makes the interpretation of the spectra more challenging. 
Here we use the Keldysh contour representation 
of effective models and nonequilibrium dynamical mean field theory to systematically study the MDCS signals of prototypical correlated lattice systems. 
By analyzing the current induced by sequences of ultrashort laser pulses we demonstrate the usefulness of MDCS as a diagnostic tool for excitation pathways and coherent processes in correlated solids. 
We also show that this technique allows to extract detailed information on the nature and evolution of photo-excited nonequilibrium states. 	
\end{abstract}

\maketitle

\section{Introduction}
	
The development of ultrashort laser pulses enabled experimental studies of the nonequilibrium properties of many-body quantum systems on their intrinsic timescales~\cite{schultze2013,giannetti2016,torre2021}. 
Time-resolved pump-probe spectroscopy has become a standard tool for exploring the response of correlated electron materials to interband charge excitations, the transient Floquet states realized under periodic driving, as well as the relaxation and thermalization processes~\cite{okamoto2010,boschini2024}. However, heating effects can obscure interesting phenomena, while nonequilibrium processes with multiple pathways can be difficult to disentangle using traditional approaches. To gain deeper insights into the dynamics of many-body systems, it is thus important to develop techniques which employ weak pulses, and which can distinguish between different excitation and relaxation pathways.  \\

Multidimensional coherent spectroscopy (MDCS) is capable of revealing nonlinear responses of atomic or molecular systems, as well as the out-of-nonequilibrium dynamics in chemical reactions and protein folding~\cite{mukamel1995,mukamel2000,hamm2011}. A more recent trend is the extension of this technique to lattice systems. A broad range of materials has been explored with pulses from the Terahertz~\cite{mahmood2021,zhang2023,barbalas2023,liu2024,liu2024a,zhang2024,gomez2024} to the optical range \cite{li2006,moody2015,yue2024}, while theoretical studies of MDCS have so far mainly focused on spin systems, few-site clusters or one-dimensional models~\cite{wan2019,li2021,gao2023,li2023,li2024,choi2020,sim2023,sim2023a,phuc2021,watanabe2024,qiang2024}.
In this article, we show how MDCS can be used to study the equilibrium and nonequilibrium properties of high dimensional strongly correlated electron systems. By simulating multi-pulse MDCS measurements for prototypical lattice models using nonequilibrium dynamical mean-field theory (NEQ-DMFT)~\cite{aoki2014}, we demonstrate that weak optical excitations can reveal the 
nature of correlated states, the interaction strengths, the relaxation of excited states, as well as coherent quantum phenomena in strongly correlated materials.  \\
	
\section{Multi-pulse set-up}
	
\begin{figure}[h]
	\centering
	\includegraphics[width=0.8\linewidth]{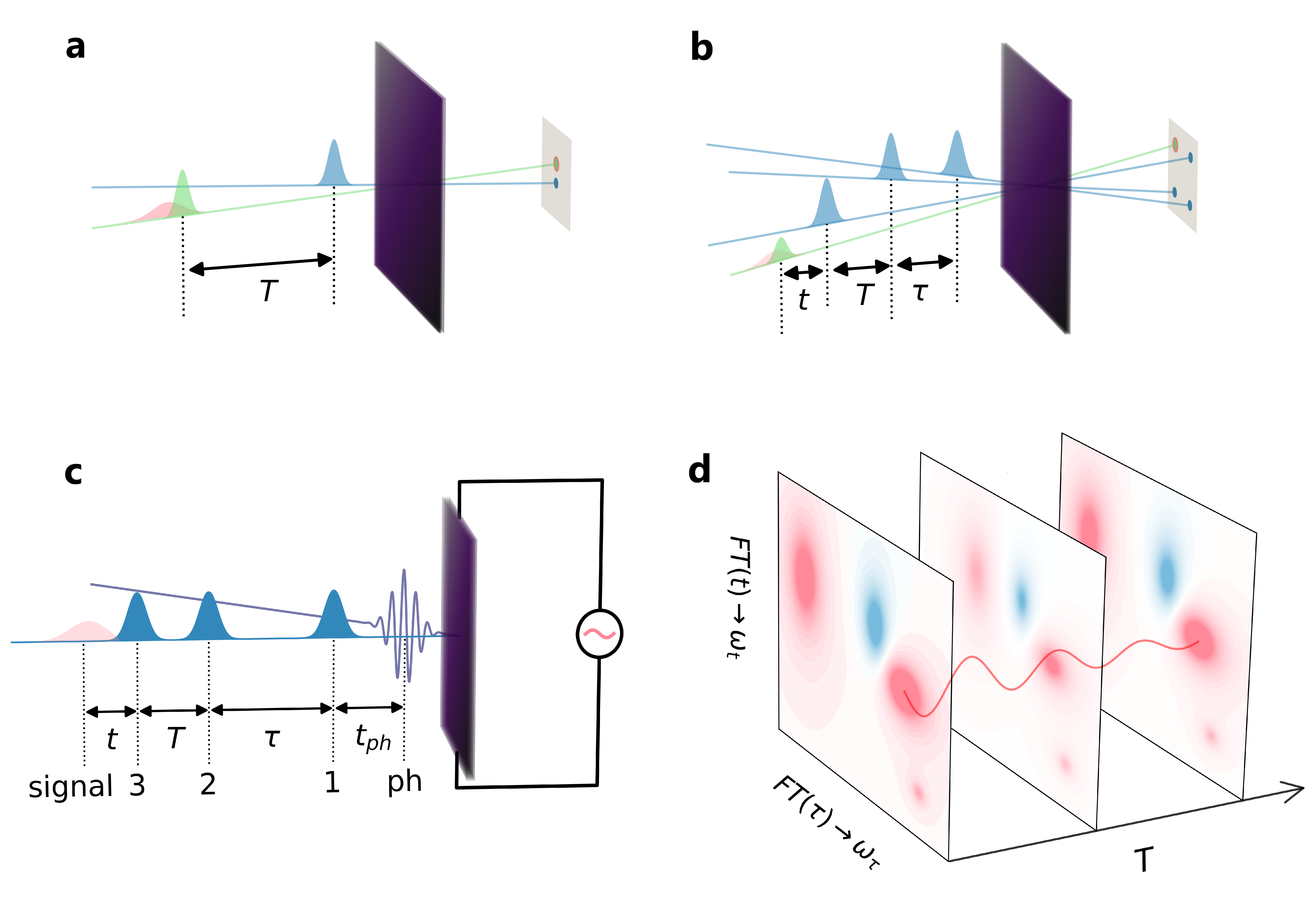}
	\caption{{\bf a}, Two-pulse setup in a pump-probe experiment. The pump (blue) and probe (green) pulses are separated by a delay time $T$. {\bf b}, Four wave mixing realization of MDCS. Four non-collinear pulses with relative time delays $\tau$, $T$ and $t$ form a box-geometry. 
		In an optical measurement, the signal (red) is heterodyned with the last pulse (green). {\bf c}, MDCS setup with three collinear pulses (blue) and optical current measurement. An additional pump excitation (purple) at time $t_\text{ph}$ before the three blue-shaded pulses can be added to photo-dope or drive the system into a nonequilibrium state. The signal (red) is collected by an electric measurement. {\bf d}, A two dimensional Fourier transformation converts the time delay $(\tau,t)$ into $(\omega_\tau,\omega_t)$. The signal intensities change as a function of the time delay $T$. }\label{fig:1-exp}
\end{figure}

Conventional transient absorption experiments involve two pulses -- a pump pulse drives the system into an excited state, while the following probe pulse is used to detect the pump-induced modifications in the sample. 
The tunable time delay between the pump and the probe pulse allows to track the 
dynamics of these changes 
(Figure~\ref{fig:1-exp}{\bf a}). While these and other pump-probe experiments provide important insights into the nonequilibrium dynamics~\cite{boschini2024}, conventional pump-probe experiments have some limitations when it comes to detecting fine and transient structures during the ultrafast evolution of the system. The typically strong pump pulse required to generate a detectable difference in the probed signal will heat the system, or even degrade the lattice structure~\cite{vidas2020}. Also, strong interband charge transfers and incoherent backgrounds 
can hide short-lived transient high energy states or coherent dynamics in strongly correlated materials~\cite{ligges2018,chen2024}. \\

In the common four-wave-mixing MDCS setup, three phase matched non-collinear pumps are applied to excite the sample in a  ``box" geometry (Figure~\ref{fig:1-exp}{\bf b}). By alternating the order of the pump sequences, different excitation pathways can be studied~\cite{branczyk2014}.
In photo-current spectroscopy (Figure~\ref{fig:1-exp}{\bf c}), the optical pumps are combined with electric current detection. 
This approach can effectively explore the low-fluence regime, suppress background signals and improve the spatial resolution~\cite{bakulin2016}.
In this work, we consider a similar protocol to simulate and analyze the two-dimensional coherent spectra (2DCS) of prototypical correlated lattice models. 
We use a fully collinear setup combining three phase-stable broadband (monocycle femtosecond) laser pulses with a direct detection of the induced current signal. The relative time delays $\tau$, $T$ and $t$ between the pulses can be precisely controlled (Figure~\ref{fig:1-exp}{\bf c}). To study the transient 2DCS signal of a photo-doped nonequilibrium system, we excite the system with an additional strong pulse which acts before or during the sequence of weak monocycle pulses (time delay $t_\text{ph}$). The third order response is obtained by subtracting the single-pulse and two-pulse contributions to the measured signal. A two-dimensional Fourier transformation within the multidimensional time delay space ($t_\text{ph}$,$\tau$,$T$,$t$) transforms ($\tau$,$t$) into ($\omega_\tau,\omega_t$). The dependence of the 2DCS signal on $t_\text{ph}$ and $T$ allows to track the evolution and decoherence of nonequilibrium states (Figure~\ref{fig:1-exp}{\bf d}).

\section{Keldysh contour analysis}
	
\begin{figure}[t]
	\centering
	\includegraphics[width=0.99\linewidth]{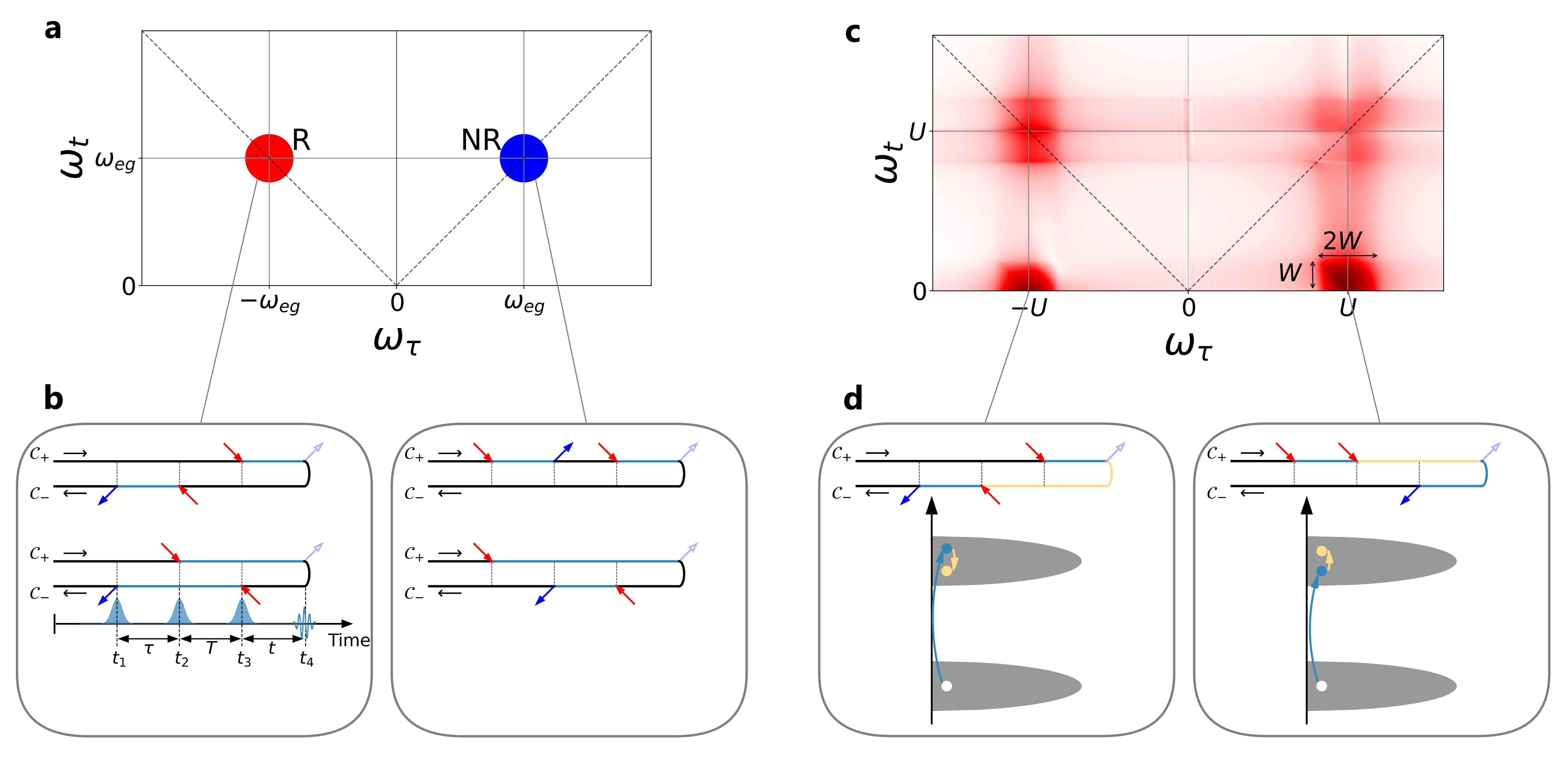}
	\caption{{\bf a}, R (red) and NR (blue) signals of a two level system with energy splitting $\omega_\text{eg}$, showing up in the $(-,+)$ and $(+,+)$ quadrant of the ($\omega_\tau$,$\omega_t$) domain. {\bf b}, Keldysh diagrams illustrating the excitation and deexcitation pathways associated with the three laser pulses (blue and red arrows). The colors of the contour segments indicate the state of the system (black: ground state, cyan: excited state). The real time intervals $\tau$, $T$ and $t$ between the three pulses and the measured signal (dashed arrow) are indicated in the bottom left subpanel. {\bf c}, Hubbard model results for local interaction $U$, 
		with on-diagonal R and NR signals due to inter-Hubbard-band excitations and additional 2Q signals associated with intra-Hubbard band excitations. {\bf d}, Keldysh diagrams and sketches of the 2Q process. Only the diagrams with the intra-Hubbard band excitation on the upper branch are shown. 
	}\label{fig:2}
\end{figure}

We simulate MDCS measurements of high-dimensional correlated lattice systems with NEQ-DMFT, which is based on the nonequilibrium Green's function  formalism~\cite{aoki2014}. 
Figure~\ref{fig:2}{\bf b} depicts the coherent evolution of the density matrix of a simple two-level system on the two-branch Keldysh contour $\mathcal{C}=\mathcal{C}_+\cup \mathcal{C}_-$. This contour representation is related to the double-sided Feynman diagrams in Liouville space~\cite{mukamel1995,hansen2012}, but more familiar to researchers with a condensed matter background. 
The arrows along the Keldysh contour mark the excitations and deexcitations of the system, while the colors of the contour segments represent the evolution of the many-body state. The three pumps induce a third-order response proportional to $S^{(3)}(\tau,T,t) \propto \langle[[[\hat{\boldsymbol{j}}(0),\hat{\boldsymbol{j}}(\tau)],\hat{\boldsymbol{j}}(\tau+T)],\hat{\boldsymbol{j}}(\tau+T+t)]\rangle$ 
(or $S^{(3)}(t_1,t_2,t_3,t_4) \propto \langle[[[\hat{\boldsymbol{j}}(t_1),\hat{\boldsymbol{j}}(t_2)],\hat{\boldsymbol{j}}(t_3)],\hat{\boldsymbol{j}}(t_4)]\rangle$]), 
where $\hat{\boldsymbol{j}}$ denotes the current operator. The three nested commutators generate eight interaction pathways, forming three groups of signals which we denote as rephasing (R, or echo), non-rephasing (NR) and two-quantum (2Q) signals, following the convention in Ref.~\cite{hamm2011}. For a detailed derivation of the nonlinear current resulting from the light-matter interaction, see supplementary material (SM) Section~\ref{sec:keldysh}. \\

In our two-level system with ground state $|g\rangle$ and excited state $|e\rangle$ (energy difference $\omega_\text{eg}$), the three pumps (solid arrows) excite $|g\rangle\to|e\rangle$ (red) or deexcite $|e\rangle\to|g\rangle$ (blue) along the Keldysh contour $\mathcal{C}$. By convention, we fix the signal emission (de-excitation $|e\rangle\to|g\rangle$, light blue arrow) at the end of $
\mathcal{C}_{+}$. 
In the upper left diagram of Figure~\ref{fig:2}{\bf b}, the pump at time $t_3$ on the $\mathcal{C}_{+}$ branch excites the system from $|g\rangle$ to
$|e\rangle$, while the signal emission at $t_4$ deexites it back to $|g\rangle$, so that the system accumulates a phase $\exp(-i\int^{t_4}_{t_3}\omega_\text{eg} ds) = \exp(-i\omega_\text{eg}t)$.
On the $\mathcal{C}_-$ branch, the excitation at time $t_2$ is followed by the de-excitation back to $|g\rangle$ at $t_1$, which yields an additional phase $\exp(-i\int^{t_1}_{t_2}\omega_\text{eg}ds) = \exp(i\omega_\text{eg}\tau)$.
The Fourier transformation of the total accumulated phase $\exp[-i\omega_\text{eg}(t-\tau)]$ produces a signal at $(-\omega_\text{eg},\omega_\text{eg})$ on the anti-diagonal in the $(\omega_\tau,\omega_t)$ domain. A similar analysis can be performed for the remaining three diagrams in Figure~\ref{fig:2}{\bf b}, and the contributions can be 
grouped into rephasing (R) and nonrephasing (NR) signals appearing in the $(-,+)$ and $(+,+)$ quadrant, respectively (Figure~\ref{fig:2}{\bf a}). 
In the presence of band broadening, the NR signals are weakened compared to the R signals (SM Section~\ref{sec:keldysh}).\\

The qualitative features of this simple two-level system are also found in the 2DCS signal of a Mott insulating single-band Hubbard model with local interaction $U$ larger than the bandwidth $W$. As shown in Figure~\ref{fig:2}{\bf c}, the spectrum for the Hubbard model features (broadened) R and NR peaks at $\omega_\tau\approx \pm U$ and $\omega_t\approx U$. In addition, this measurement produces 2Q signals at  $\omega_\tau\approx \pm U$ and $\omega_t\lesssim W$, representing the excitation and deexcitation processes within the Hubbard bands. 2Q signals correspond to diagrams where the first two pumps along the real time axis interact with the system on the same branch (Figure~\ref{fig:2}{\bf d}). A weaker (stronger) first excitation $\omega_\tau$, populating states at the lower (upper) edge of the upper Hubbard band, allows for larger in-band excitations (deexcitations).
Consistent with this, a close inspection of the intensities of the 2Q signals in Figure~\ref{fig:2}{\bf c} shows that  the largest signal appears along a line with negative slope in the $(\omega_\tau,\omega_t)$ plane. This effect is even more pronounced in a system with a larger bandwidth $W$, as shown in the  Figure~\ref{fig:ex}. \\

In lattice systems with multiple orbitals per site or unit cell,
the more complex electronic structure results in numerous excitation and relaxation pathways, coherences, and intermediate states with different characteristic lifetimes, as we will demonstrate with the following examples. \\

\begin{figure}[t]
	\centering
	\includegraphics[width=0.99\linewidth]{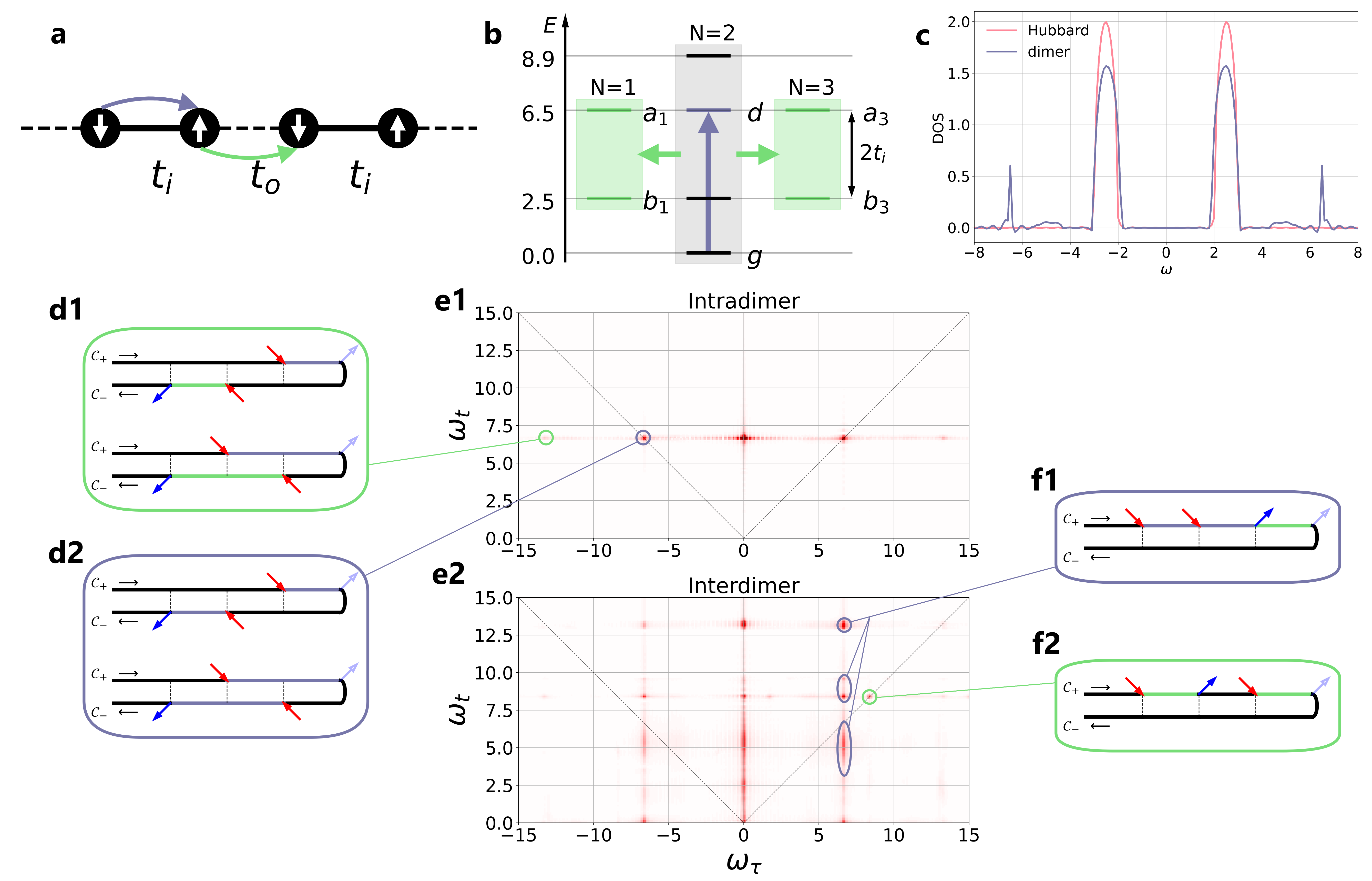}
	\caption{ {\bf a}, Sketch of the correlated dimer model. The solid bonds with hopping $t_i$ form dimers, while the dashed bonds with hopping $t_o$ connect adjacent dimers.
		Excitations within the dimer (purple arrow) create a doublon/holon state on the dimer. 
		{\bf b}, Charge excitations between dimers (green arrow) create triply occupied and singly occupied dimers, with additional electrons or holons in the antibonding or bonding states. {\bf c}, Interacting density of states of the Mott insulating single band Hubbard model (red) and the dimerized correlated band insulator (blue). {\bf e}, 2DCS signal from the intradimer ({\bf e1}) and interdimer ({\bf e2}) current for the correlated dimer model. We use different color ranges since the intensity of the intradimer signal is larger than that of the interdimer signal. {\bf d},{\bf f}, Keldysh diagrams for the indicated features in the interdimer 2DCS signal. 
	}\label{fig:3}
\end{figure}
	
\section{Revealing the nature of a correlated insulator}

Scanning tunneling microscopy, optical conductivity and photoemission spectroscopy measurements can reveal the density of states (DOS), gap sizes and band structures of strongly correlated materials. However, in some systems, like the layered polaronic insulator $1T$-TaS$_2$ or the Peierls distorted system VO$_2$, dimerizations of atoms or modulations in the stacking arrangement lead to a nontrivial interplay of band-insulating and Mott insulating characteristics~\cite{biermann2005,ritschel2018,petocchi2023,petocchi2022,chen2024}.  
Determining the nature of an insulator from the density of states alone is usually not possible. Here we demonstrate that 2DCS measurements yield clearly distinct signals for a Mott insulator and a correlated band insulator. \\

The 2DCS signal of a pure Mott insulator (single-band Hubbard model with interaction $U=5$ and bandwidth $W=4v$, $v=0.25$) has been shown in Figure~\ref{fig:2}{\bf c}, and the corresponding single-particle spectral function is plotted by the red line in Figure~\ref{fig:3}{\bf c}. The spectral function features a gap of size $U-W=4$. 
A very similar spectral function can be obtained in a correlated dimer setup inspired by the bilayer stacking arrangement in 1$T$-TaS$_2$~\cite{lee2019}, see blue line in Figure~\ref{fig:3}{\bf c} (the realistic material corresponds to a unit of energy of approximately $0.1$ eV). In this system, the charge gap is a correlated hybridization gap~\cite{petocchi2022,petocchi2023}, controlled by excitations between bonding/antibonding states. We use a cluster DMFT setup with intradimer (interdimer) hoppings $t_i=2$ ($t_o = 0.45$) and onsite interaction $U=4$. The resulting DOS has two main peaks separated by $\sqrt{U^2+16t_i^2}-U=4.9$ and a similar bandwidth as in the single band Hubbard model. \\ 

The 2DCS signal for the correlated band insulator, shown in Figure~\ref{fig:3}{\bf e}, differs significantly from that of the Mott insulator. We show two spectra corresponding to the intra-dimer (Figure~\ref{fig:3}{\bf e1}) and inter-dimer (Figure~\ref{fig:3}{\bf e2}) currents. The intra-dimer signal has dominant R and NR peaks at $(\omega_\tau,\omega_t) = (\pm 6.7, 6.7)$.  The energy $\Delta E=(\sqrt{U^2+16t_i^2}+U)/2= 6.5$ corresponds to the creation of a doublon/holon pair within a dimer (purple arrows in Figs.~\ref{fig:3}{\bf a},{\bf b}), and the increase to $6.7$ is the effect of the small inter-dimer hopping $t_o$. The Keldysh diagrams for the R peak are shown in Figure~\ref{fig:3}{\bf d2}. In addition, there are satellites at $(\omega_\tau,\omega_t) = (\pm 13.2, 6.7)$ which involve inter-dimer charge excitations from/to the ground state (see green arrows in Figure~\ref{fig:3}{\bf a},{\bf b} and the Keldysh diagrams in Figure~\ref{fig:3}{\bf d1}). \\

The inter-dimer 2DCS signal has a rich structure which reveals additional excitation and deexcitation processes. Let us focus on the column of peaks at $\omega_\tau = 6.7$ in the NR quadrant. This excitation energy indicates a coupling of the inter-dimer current to the intra-dimer doublon/holon processes. 
While the state after the first two pulses is still dominated by $N=2$ dimers, inter-dimer hopping then produces charge excitations $(N=2,N=2)\rightarrow (N=1,N=3)$ to bonding/antibonding states with energy splittings of roughly $\Delta E=5$, 9, 13 relative to the ground state (Figure~\ref{fig:3}{\bf b},{\bf f1}). The three peaks marked by the purple ovals in Figure~\ref{fig:3}{\bf e2} represent inter-dimer singlon-triplon annihilation processes which bring the system back to the ground state. Here, the middle peak is split into two subpeaks at $\omega_t=8.4$ and $9.6$ by the effect of the inter-dimer hopping $t_o$. The signal at $\omega_t<0.5$ reflects intra-band excitation/deexcitation processes. \\ 

There are additional peaks in this quadrant which have a simple interpretation. For example the peak at $(\omega_\tau,\omega_t)=(8.4,8.4)$ corresponds to the sequence of inter-dimer charge excitation processes sketched in Figure~\ref{fig:3}{\bf f2}, which involves direct charge excitations to/from the intermediate energy manifold of the singlon-triplon pairs $(N=1,N=3)$. Similarly, the signal at $(\omega_\tau,\omega_t)=(13,13)$
involves the high-energy manifold. The signals with $\omega_\tau \approx 0$ in Figure~\ref{fig:3}{\bf e2} (Figure~\ref{fig:3}{\bf e1}) must correspond to intra-dimer (inter-dimer)  excitation processes which do not directly couple to the inter-dimer (intra-dimer) current. \\
	
\section {Coherence in multi-orbital systems}
	
Most strongly correlated materials host multiple active orbitals. 
The nontrivial interplay of bandwidths, crystal field splittings, Hubbard and Hund interactions complicates the interpretation of experimental results and the choice of appropriate parameters for numerical calculations. Here, we show that 2DCS signals of multi-orbital lattice systems allow to extract the interaction parameters and the decoherence times of excited states.  \\

\begin{figure}[t]
	\centering
	\includegraphics[width=\linewidth]{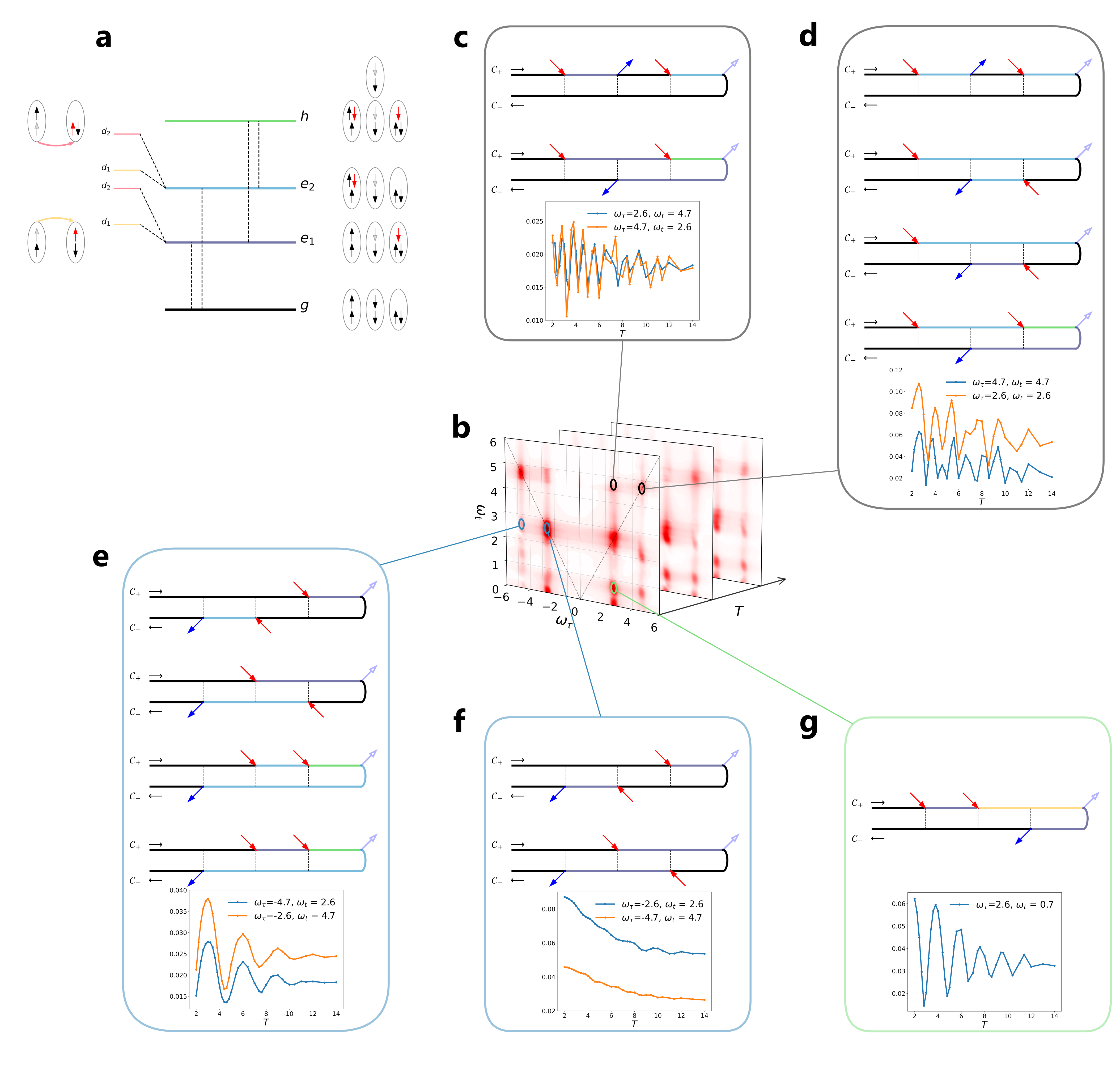}
	\caption{
		{\bf a}, Relevant configurations for the spin crossover model: ground state $g$, excited state $e_1$, $e_2$, doubly excited state $d_1$,$d_2$, possible high energy state  $h$. {\bf b},  2DCS signal for a given waiting time $T$. {\bf c}-{\bf g}, Signal intensity as a function of $T$ and corresponding contour diagrams corresponding to the blue lines for the NR off-diagonal cross peak ({\bf c}), NR diagonal peak ({\bf d}), R off-diagonal cross peak  ({\bf e}), R diagonal peak ({\bf f}) and 2Q signal ({\bf g}).
		Orange lines share the properties of the blue lines, but correspond to a different excitation process ($e_1$ or $e_2$).
	}
	\label{fig:4}
\end{figure}

We study a two-orbital Hubbard model with Coulomb interaction $U$, Hund coupling $J$, crystal field splitting $\Delta$ and orbital-diagonal hopping $v$,
\begin{equation}\label{eq:2orb}
	\hat{H} =  -\sum_{\langle ij\rangle,a,\sigma} v\hat{c}^\dagger_{ia\sigma}\hat{c}_{ja\sigma}+ \sum_{i,a} U \hat{n}_{ia\uparrow} \hat{n}_{ia\downarrow} + \sum_{i,a>b} (U-2J) \hat{n}_{ia} \hat{n}_{ib} -\sum_{i,a>b,\sigma} J\hat{n}_{ia\sigma} \hat{n}_{ib\sigma} +\sum_{i,a>b}\Delta (n_{ia}-n_{ib})-\mu N.
\end{equation}
Here, $n_{ia\sigma}$ denotes the occupation of orbital $a$ at site $i$ with electrons of spin $\sigma$, $n_{ia}=n_{ia\uparrow}+n_{ia\downarrow}$, and the sum in the first term is over nearest-neighbor sites. We solve the lattice model on an infinite-dimensional Bethe lattice, with a rescaled hopping parameter $v=0.1$ corresponding to a bandwidth $W=4v=0.4$ (same for both orbitals). 
The interactions are chosen as $U=4$ and $J=0.7$. 
The procedure described in Ref.~\onlinecite{werner2017} is used for the simulation of electric field pulses. 
We place ourselves close to the spin-state transition between the low-spin (LS) and high-spin (HS) insulating phases~\cite{werner2007} in a system with $\Delta= 2.35$
and choose the chemical potential $\mu$ corresponding to half filling. \\

For the analysis of the data, it is useful to consider a simple few-level scheme which captures the relevant quasi-local processes. 
In Figure~\ref{fig:4}{\bf a}, we sketch the ground state manifold ($g$), which contains all the nearly degenerate high-spin and low-spin doublon states, 
excited states ($e_1$, $e_2$) with excitation energies $\omega_{e_{1}}=U-2J$ and $\omega_{e_{2}}=U+J$, second-order excitations ($d_1$, $d_2$) with an additional cost of $J$ and $3J$, and a high energy state $h$. The two possible direct excitations to $e_1$ and $e_2$ yield the diagonal peaks at $(\omega_\tau,\omega_t)=(\pm 4.7,4.7)$ and $(\pm 2.6,2.6)$ for the R/NR pathways (Figure~\ref{fig:4}{\bf b}).  
The cross peaks at $(\pm 4.7,2.6)$ and  $(\pm 2.6, 4.7)$ demonstrate the coupling of $e_1$ and $e_2$ through the ground states $g$ or the high energy state $h$.\\

In Figure~\ref{fig:4}{\bf c}-{\bf g}, we show R, NR and 2Q signal intensities as a function of $T$, together with the corresponding Keldysh contour diagrams. The R signals have clean features.
The off-diagonal cross peaks (Figure~\ref{fig:4}{\bf e}) oscillate at the frequency of $\omega_{e1e2}=2.1$, since two of the diagrams are in a superposition state during $T$. The diagonal signals (Figure~\ref{fig:4}{\bf f}) show a nonoscillating decay, since the system is in a population state $|e\rangle\langle e|$ or $|g\rangle\langle g|$ during the interval $T$. In contrast, the NR signals oscillate with multiple frequency components (Figure~\ref{fig:4}{\bf c},{\bf d}), due to their nonrephasing nature. \\

The feature located at $(2.6, 0.7)$ (Figure~\ref{fig:4}{\bf g}) is a 2Q signal. As revealed by the Keldysh contour analysis, the first excited state $e_1$ undergoes an additional Hund excitation involving the formation of an interorbital LS state ($d_1$) with $\omega_{e_1d_1}=J=0.7$ (yellow arrow on the left side of Figure~\ref{fig:4}{\bf a}). During the $T$ interval, the local state is a superposition state $|d_1 \rangle\langle g |$, whose intensity oscillates at the frequency $E_{d_1}- E_g = 3.3$ (Figure~\ref{fig:4}{\bf g}). The signals with energies below $\omega_t<0.4$ represent excitation/deexcitation processes within the band. By combining such analyses of the dominant signals in the  2DCS spectrum, the interaction parameters $U$ and $J$ of the simulated model can be determined. \\

In the strongly correlated regime $U/W\gg 1$, the life-time $T_p$ of photo-excited charge carriers, which scales $\propto \exp[U/W]$ \cite{sensarma2010}, is much longer than the time range of our simulations. While this life-time has been studied both experimentally and theoretically \cite{strohmaier2010,eckstein2011}, the decoherence time $T_d$ of the charge carriers has not been accessible before. The decay of the intensities of the 2DCS signals during the waiting time $T$ allow us to extract $T_d$. In SM~Tabel~\ref{table1}, we list the coherence times obtained from fits to the NEQ-DMFT data. These times are comparable to the inverse hopping time $\hbar/v\equiv v^{-1}=10$.
Note that the signals decay to a nonzero constant, representing the populated state with a much longer lifetime $T_p$.\\

\section{Transient 2DCS of a photo-doped insulator}
	
2DCS can also be used to study out-of-equilibrium systems, for example, molecular systems after a strong (actinic) laser excitation~\cite{hamm2021}. Here, we demonstrate the power of the method by considering a photo-doped Mott insulator.  The additional photo-doping pulse is applied to the system a time $t_\text{ph}$ before the three phase stable pumps of the 2DCS measurement (Figure~\ref{fig:1-exp}{\bf c}). \\ 
	
\begin{figure}[t]
	\centering
	\includegraphics[width=0.87\linewidth]{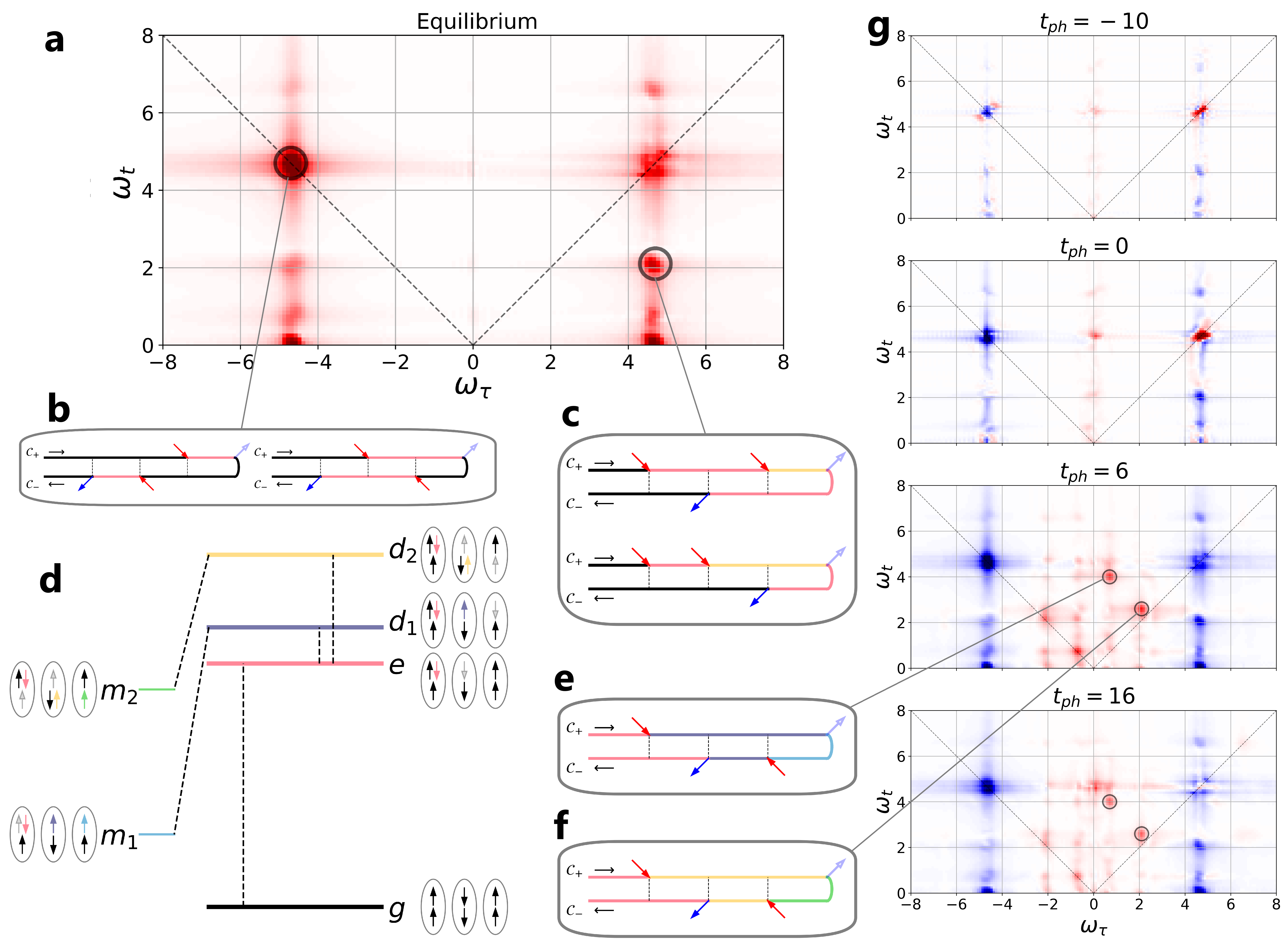}
	\caption{
		{\bf a}, Equilibrium 2DCS signal for the high-spin insulator.
		{\bf b},{\bf c}, Keldysh diagrams of prominent signals. 
		{\bf d}, Spin configurations for the ground state $g$, excited state $e$, doubly excited states $d_1$,$d_2$ and metastable states $m_1$,$m_2$.
		{\bf e},{\bf f}, Keldysh diagrams of relevant signals in the difference spectrum. 
		{\bf g}, 2DCS difference spectra for a photo-doped system at the indicated pump delay times $t_\text{ph}=-10$, $0$, $6$ and $16$ relative to the 2DCS measurement. 
	}\label{fig:5}
\end{figure}

We consider the half-filled two-orbital Hubbard model (Eq.~\eqref{eq:2orb}) without crystal field splitting ($\Delta=0$). The other parameters ($U=4$, $J=0.7$, $v=0.1$) are the same as in the previous example. The ground state $g$ corresponds to the HS insulator (Figure~\ref{fig:5}{\bf d}). Focusing on nearest-neighbor hopping processes, the photo-doping pulse can only generate a single type of excitation, from high-spin doublons to a singlon-triplon pair (excited state $e$), at an energy cost of $U+J$. The 2DCS pulse sequence is applied before, during and after the photo-doping pulse to trace the excitation process and the relaxation of the system into a transient nonequilibrium state. As a reference, Figure~\ref{fig:5}{\bf a} shows the equilibrium 2DCS signal. In addition to the diagonal peaks at $(\omega_\tau,\omega_t)=(\pm 4.7, 4.7)$, there exist cross peaks located at $\omega_t = 2.1, 0.7$, representing excited state transitions to doubly excited states $d_1$ or $d_2$  (Figure~\ref{fig:5}{\bf d}) at the cost of $J$ or $3J$. The feature at $(\omega_\tau,\omega_t)=(\pm 4.7, 6.6)$ results from an excited state with energy $2U-2J$ relative to $e$. Such a state can be reached for example by a hop of the middle spin down in configuration $e$ to the right. 
	In addition, there is a signal close to $\omega_t=0$, representing intra-band excitation and deexcitation processes. \\
	
In Figure~\ref{fig:5}{\bf g}, we plot difference-2DCS spectra (with the equilibrium signal subtracted) for the indicated delay times $t_\text{ph}$ between the photo-doping pulse and the sequence of pump pulses. The signals located at $\omega_\tau = \pm 4.7$ become negative, revealing the bleaching effect of the photo-doping pulse, which transfers population from the $g$ state to the $e$ state. The positive signals indicate the third order response of $e$. In particular, the signals located at the columns $\omega_\tau = \pm J=\pm 0.7$ ($\pm 3J=\pm 2.1$) represent the injection of the corresponding energies by the first pump (gray circles). Such excitations become possible, because the photo-doped nonequilibrium state contains a significant population of singly and triply occupied sites (see excitations from $e$ to $d_1$ and $e$ to $d_2$ in Figure~\ref{fig:5}{\bf d}). Dominant signals associated with such processes appear at emission energy $\omega_t=U=4$ and $\omega_t=U-2J=2.6$, see Keldysh diagrams in Figure~\ref{fig:5}{\bf e},{\bf f}). The signal at $\omega_\tau\sim 0$ and $\omega_t=4.7$ reveals the metallic nature of the photo-induced transient state (black box). \\
	
\section{Summary}
	
We showed how the 2DCS signals of correlated lattice models can be calculated with NEQ-DMFT and interpreted in terms of Keldysh diagrams for few-level systems representing quasi-local processes. These signals reveal information on the nature of the ground state, the excitation or relaxation processes and the coherence times, which are difficult to obtain from standard absorption, conductivity or photoemission spectroscopy measurements. For example, the 2DCS spectra can clearly distinguish a simple Mott insulator from a correlated band insulator with dimerization, because the more complex energy level structure of the latter system leads to a much richer 2DCS signal. Similarly, the 2DCS spectra of multi-orbital Mott systems clearly reveal the relevant interaction parameters $U$ and $J$, and the activation of new degrees of freedom (like mobile singlons and triplons) in a photo-doped Mott state. \\

2DCS can measure nonlinear responses with a series of weak probe pulses, which do not significantly perturb the probed many-body state. In the present study, we have focused on a collinear setup within a Bethe-lattice-type construction that does not exploit the polarization of the light. In more realistic simulations, additional information could be obtained by varying the geometry, polarization vectors, and relative phases of the pump pulses.     

\acknowledgements
We thank M. Eckstein and F. Petocchi for helpful discussions. J.C. also thanks Z. Li and Y. Wan for motivating discussions.
This work was supported by the Swiss National Science Foundation through the Research Unit QUAST of Deutsche Foschungsgemeinschaft (FOR5249), and Grant No.~200021-196966.   The calculations were performed on the Beo05 cluster at the University of Fribourg.\\

\bibliographystyle{unsrt}
\bibliography{ref}

\newpage
\begin{center}
{\bf MULTIDIMENSIONAL COHERENT SPECTROSCOPY OF CORRELATED LATTICE SYSTEMS\\
SUPPLEMENTARY
MATERIAL}
\end{center}

\appendix
\section{Fits of the interval $T$ dependence of the NEQ-DMFT data}
\begin{table}[!h]
\begin{center}
	\begin{tabular}{|>{\centering}p{5mm} | >{\centering}p{2cm} | >{\centering}p{2cm} | c | } 
		\hline
		&peak location  & oscillation  & decoherence  \\ 
		&($\omega_\tau$,$\omega_t$) & frequency $\Omega$ & time $\mathcal{T}$ \\ \hline\hline
		\multirow{4}*{R}
		&$(-2.6, 2.6)$ & - & 4.1  \\  \cline{2-4}
		&$(-4.7, 4.7)$ & - & 4.9  \\  \cline{2-4}
		&$(-2.6, 4.7)$ & 2.03 & 3.4  \\  \cline{2-4}
		&$(-4.7, 2.6)$ & 2.08 & 3.9 \\  \hline\hline
		\multirow{4}*{NR}
		&$(2.6, 2.6)$ & 2.6 & -  \\ \cline{2-4}
		&$(4.7, 4.7)$ & 4.6 & -  \\ \cline{2-4}
		&$(2.6, 4.7)$ & 7.1 & -  \\ \cline{2-4}
		&$(4.7, 2.6)$ & 7.0 & -  \\ \hline\hline
		\multirow{4}*{2Q}
		&	$(2.6, 2.1)$ & 4.7 & 5.5  \\ \cline{2-4}
		&	$(2.6, 0.7)$ & 3.3 & 5.5  \\ \cline{2-4}
		&	$(4.7, 2.1)$ & 6.8 & 7.6  \\ \cline{2-4}
		&	$(4.7, 0.7)$ & 5.4 & 7.5  \\ \hline
	\end{tabular}
	\caption{Oscillation frequency $\Omega$ and decoherence time $\mathcal{T}$ extracted from the fit to the $T$-dependence of the R (top), NR (middle) and 2Q (bottom) signals shown in Figure~\ref{fig:4}.
		We used the fitting function $I(T) = |(a\cos(\Omega T+\phi) + b)e^{-T/\mathcal{T}} +c|$ which captures constant, exponentially decaying and coherently oscillating components, representing the ground state, populated excited state and superposition state during $T$.}\label{table1}
\end{center}
\end{table}

\clearpage

\section{Decoherence of a Mott insulator in the time domain}
	
\begin{figure}[h]
	\centering
	\includegraphics[width=0.8\linewidth]{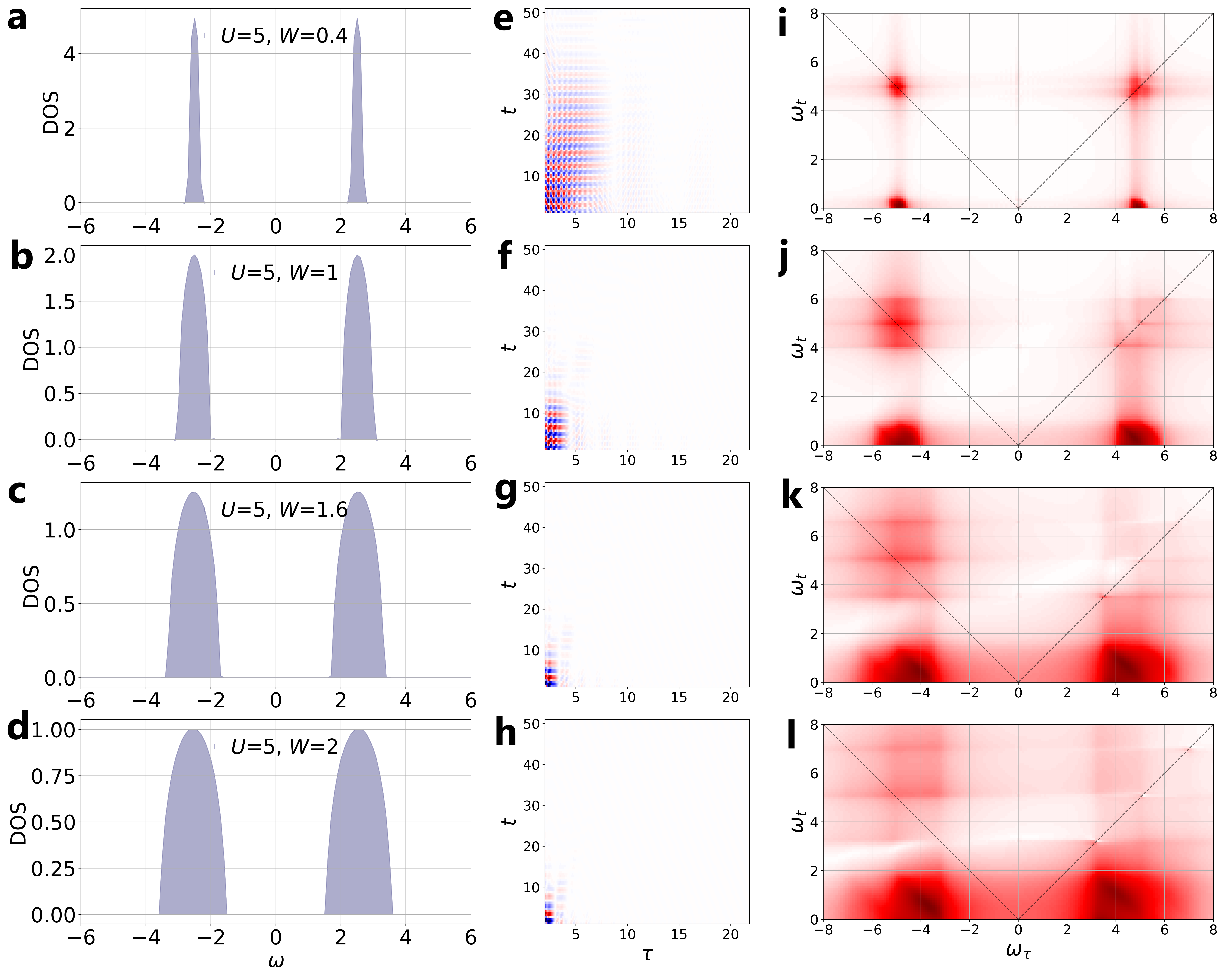}
	\caption{{\bf a}-{\bf d}, Density of states (DOS) for $U=5$ and indicated bandwidths $W$. {\bf e}-{\bf h}, 2DCS signal in the time domain for waiting time $T=4$. The decoherence time gets shorter as the bandwidth $W$ increases. {\bf i}-{\bf l}, 2DCS signal after the Fourier transform from $(t,\tau)$ to $(\omega_t,\omega_\tau)$. The 2Q signal becomes more prominent with increasing $W$, revealing more clearly the intra-Hubbard band dispersion. Both the R and NR signals become blurred as $W$ increases, while the relative intensity of the R signal increases. } 
	\label{fig:ex}
\end{figure}

\section{Nonequilibrium Dynamical Mean-Field Theory}
Given a time dependent Hamiltonian $\hat{H}(t)$, a contour ordered correlation function on the three-leg Kadanoff-Baym contour $\mathcal{C}$ is defined as \cite{aoki2014} 
\begin{equation}
	C_{AB}(t,t')=\langle \mathcal{T}_\mathcal{C} \hat{A}(t)\hat{B}(t') \rangle = \frac{1}{Z}\tr \Big[ \mathcal{T}_\mathcal{C} e^{ - i\int_\mathcal{C} d\bar t H(\bar t) } \hat{A}(t)\hat{B}(t')\Big].
\end{equation}
Here, $\mathcal{T}_\mathcal{C}$ denotes the contour time ordering operator. The integral over the Matsubara-branch of the three-leg contour, divided by the partition sum $Z$, corresponds to the density matrix of the initial state. 

Using a cavity construction, dynamical mean field theory (DMFT) maps the lattice problem to an effective impurity model with a self-consistently determined bath, represented in action form by the hybridization function $\Delta(t,t')$~\cite{georges1996}. The solution of the impurity model yields the impurity Green's function, 
\begin{equation}
	G(t,t')=-i\frac{1}{Z}\tr \Big[\mathcal{T}_{\mathcal C} e^{ - i\int_\mathcal{C} d\bar t\hat{H}_\text{loc}(\bar t)-i\int_\mathcal{C} d\bar t d\bar t' c^\dagger(\bar t)\Delta(\bar t,\bar t')c(\bar t')}\hat{c}^\dagger(t)\hat{c}(t') \Big],
\end{equation}
where for simplicity, we omitted spin and orbital indices.
The DMFT self-consistency loop defines $\Delta$ in such a way that the impurity Green's function becomes identical to the local lattice Green's function. In the case of the infinitely conneced Bethe lattice, one can derive a direct relation between these two functions: $\Delta(t,t') = h(t')G(t,t')h(t)$, where $h(t)$ is the (properly renormalized) nearest-neighbor hopping~\cite{georges1996}. We also use such a relation to approximately treat other lattices, like the dimerized chain~\cite{petocchi2023}. The non-cross approximation (NCA) is used as a nonequilibrium impurity solver~\cite{eckstein2010}.

The electric current in the lattice model is given by
\begin{equation}	{\boldsymbol{j}}(t) = \langle \hat{\boldsymbol{j}}(t) \rangle = i e\sum_{i, j, s} h_{i j} \boldsymbol{R}_{i j} \langle \hat{c}_{j s}^{\dagger}(t) \hat{c}_{i s}(t)\rangle = e\sum_{i, j, s} h_{i j} \boldsymbol{R}_{i j} G^<_{ij,s}(t,t), 
\end{equation} 
where $\boldsymbol{R}_{i j}$ is the vector connecting site $i$ and $j$ and $h_{ij}$ the corresponding hopping amplitude. It can be directly calculated when the sites $i$ and $j$ are included in the DMFT impurity problem (e. g. for the dimer).
To obtain the the electric current between impurity ($i$) and bath ($j$) sites, the nonequilibrium DMFT estimate is~\cite{lysne2020}
\begin{equation} 
	\boldsymbol{j}(t) =  -i e \sum_{ij}\boldsymbol{R}_{i j} [G_i\ast \Delta_j]^<(t,t), 
\end{equation} 
where $\ast$ is a convolution on $\mathcal{C}$ and $\Delta_j$ is the hybridization function associated with the hopping to $j$.

\section{Nonlinear current}
The current calculated above is nonperturbative and contains the response to the excitation to all orders. To extract the second (2nd) or third order (3rd) contribution to the electric current, we subtract the lower order currents. Applying the three incoming weak laser pulses $A$, $B$, $C$ at time $0$, $\tau$, and $\tau+T$, we thus obtain the second and third order current responses by calculating
\begin{equation}
	\begin{aligned}
		J^{AB}_{2nd}(\tau,t)\equiv& \, J^{AB}(t+\tau)-J^A(t+\tau)-J^B(t+\tau),\\
		J^{ABC}_{3rd}(\tau,t) \equiv& \, J^{ABC}(t+\tau+T)\\ 
		&-J^{AB}_{2nd}(t+\tau+T)-J^{AC}_{2nd}(t+\tau+T)-J^{BC}_{2nd}(t+\tau+T)\\
		&-J^A(t+\tau+T)-J^B(t+\tau+T)-J^C(t+\tau+T), 
	\end{aligned}
\end{equation}
where $J^{X}(\mathscr{T})$ denotes for the current measured at time $\mathscr{T}$ in presence of the pulse combinations $X=A$, $B$, $C$, $AB$, $BC$, and $ABC$. A 2D Fourier transformation $t\rightarrow \omega_t$, $\tau\rightarrow \omega_\tau$ then yields
$J^{AB}_{2nd}(\omega_\tau,\omega_t)$ and $J^{ABC}_{3rd}(\omega_\tau,\omega_t)$. 
Finally, we take the absolute value 
$I^{ABC}_{3rd}(\omega_\tau,\omega_t) \equiv |J^{ABC}_{3rd}(\omega_\tau,\omega_t)|$ for the analysis of the spectra.\\

\section{Electric field pulses}

In a gauge with pure vector potential, the effect of an electric field is to dress the hopping amplitudes with a complex Peierls phase $\phi_{ij}(t) = \int_i^j \vec{A}(\vec{r},t) d\vec{r}$, 
\begin{equation}
	h_{ij}(t) = h_{ij}e^{i\phi_{ij}(t)}.
\end{equation}
The vector potential itself is the time integral of the electric field of the laser, $\vec{A}(\vec{r},t) =  \int_0^t dt' \vec{E}(\vec{r},t')$. To generate a broad band excitation for 2DCS, we use a broadened delta function in time as the vector potential $A(t) = A_0 \frac{1}{\sqrt{2\pi}\sigma}e^{-\frac{(t-t_0)^2}{2\sigma^2}}$. 
Specifically, we use $\sigma=0.2$ for the single-orbital Hubbard model and correlated dimer model, and $\sigma=0.25$ for the two-orbital Hubbard model. The corresponding power spectrum of the electric field is $|E(\omega)|^2 = A_0 \omega e^{-\sigma^2\omega^2}$.

For the photo-doping in the transient 2DCS simulations, we employ a strong multi-cycle electric field pulse 
\begin{equation}
	E(t) = E_0 \frac{1}{\sqrt{2\pi}\sigma}e^{-\frac{(t-t_0)^2}{2\sigma^2}}\sin(\omega_0(t-t_0)).
\end{equation}
Here, the pulse frequency $\omega_0$ (comparable to the gap size) is chosen to maximize absorption and the width of the envelope $\sigma$ is large enough to acommodate multiple cycles.

\section{High order Kubo formula}

We consider a time dependent Hamiltonian in an interaction picture, where the full time dependent $\hat{H}(t)$ is split into a time independent part $\hat{H}_0$ and a time dependent perturbation $\hat{H}'$, i.e., $\hat{H}(t) = \hat{H}_0 + \hat{H}'(t)$. Defining $|\psi_I(t)\rangle = e^{i\hat{H}_0t}|\psi_S(t)\rangle$ and  $\hat{H}'_{I}(t) = e^{i\hat{H}_0t}\hat{H}'(t)e^{-i\hat{H}_0t}$, one obtains the Liouville-von Neumann equation of motion for the density matrix 
\begin{equation}
	\frac{\mathrm{d}}{\mathrm{d}t}\hat{\rho}_I(t) = -i[\hat{H}'_I(t),\hat{\rho}_I(t)].
\end{equation}
The solution is $$\hat \rho_I(t_1) = \mathscr{U}^\dagger_I\left(t_1, t_0\right) \hat \rho_I(t_0) \mathscr{U}_I\left(t_1, t_0\right),$$ with the time evolution operator $\mathscr{U}_I\left(t_1, t_0\right) = \mathcal{T} \exp \left[-i \int_{t_0}^t d \bar{t} H'_I(\bar{t})\right]$. The density matrix $\hat{\rho}_I$ can be expended in powers of $\hat{H}'$ as
$\hat{\rho}_I(t) = \sum_i\hat{\rho}^{(i)}_I(t)$~\cite{nozieres1997},
\begin{equation}\label{eq:rho-order}
	\rho^{(i)}_I(t) =  (-i)^n\int dt_i\int dt_{i-1}\ldots \int dt_1[H'_I(t_i),[H'_I(t_{i-1})\dots,[H'_I(t_1),\rho(0)]\dots]].
\end{equation}

For an operator $\hat{\boldsymbol{j}}$, the corresponding observable $\boldsymbol{j}(t)$ can also be expended into series $\boldsymbol{j}(t) = \sum_i \boldsymbol{j}^{(i)}(t)$, with
\begin{equation}\label{eq:j-order}
	\boldsymbol{j}^{(i)}(t) = \tr\left(\hat{\boldsymbol{j}}_I(t)\hat{\rho}^{(i)}(t)\right).
\end{equation}

\section{High order electric current in correlated lattice models}
In the velocity gauge, the vector potential ${\bf A}(t)$ satisfies $\vec{\nabla}\cdot {\bf A}= 0$, representing a transverse electromagnetic field. The electrical current is obtained by $\hat{\boldsymbol{j}}(t) = -\delta \hat{H}(t)/\delta{\boldsymbol{A}}$~\cite{lenarcic2014}. Specifically, for an interacting tight-binding model,
\begin{equation}
	\begin{aligned}
		\hat{H}[\boldsymbol{A}(t)]&=-\sum_{i, j, s} h_{i j} \exp \left[i e \boldsymbol{A}(t) \cdot \boldsymbol{R}_{i j}\right] \hat{c}_{j s}^{\dagger} \hat{c}_{i s}+\hat{H}_{\mathrm{int}},\\
		\hat{\boldsymbol{j}}(t)=-\frac{\delta \hat{H}}{\delta \boldsymbol{A}}(t)&=ie\sum_{i, j, s} h_{i j}\boldsymbol{R}_{i j} \exp \left[i e \boldsymbol{A}(t) \cdot \boldsymbol{R}_{i j}\right] \hat{c}_{j s}^{\dagger} \hat{c}_{i s}.
	\end{aligned}
\end{equation}
Combining the equations above, the $n$-th order current becomes
\begin{equation}\label{eq:current_order}
	\boldsymbol{j}^{(i)}(t) = \tr\left(\hat{\boldsymbol{j}}_I(t)\rho^{(i)}(t)\right) = \tr\left(\hat{\boldsymbol{j}}_I(t)(-i)^n\int dt_i\int dt_{i-1}\int dt_1[H'_I(t_i),[H'_I(t_{i-1}),[\dots,[H'_I(t_1),\rho(0)]\dots]]]\right).
\end{equation}
The perturbation term
$$\hat{H}'_I(t) = e^{i\hat{H}_0t}\left\{-\sum_{i, j, s} h_{i j} \left(1-\exp \left[i e \boldsymbol{A}(t) \cdot \boldsymbol{R}_{i j}\right] \right)\hat{c}_{j s}^{\dagger} \hat{c}_{i s})\right\}e^{-i\hat{H}_0t}$$
vanishes when $\boldsymbol{A}(t)=0$, 
and $\hat{H}_0$ is the Hamiltonian without laser field,
\begin{equation}
	\hat{H}_0=-\sum_{i, j, s} h_{i j} \hat{c}_{j s}^{\dagger} \hat{c}_{i s}+\hat{H}_{\mathrm{int}}.
\end{equation}
Expanding $H'_I(t)$ into a series of the vector potential ${\bf A}(t)$, we get
\begin{equation}
	H'_I(t)=e^{iH_0t}\left\{-
	\hat{\boldsymbol{j}}_0 \mathbf{A}(t)+\frac{1}{2}  \hat{\tau} \mathbf{A}(t) \otimes  \mathbf{A}(t) + \dots\right\}e^{-iH_0t},
\end{equation}
with the linear current operator $\hat{\boldsymbol{j}}_0$ and stress tensor operator $\hat \tau$ are given by
\begin{equation}
	\hat{\boldsymbol{j}}_0=ie \sum_{i, j, s} h_{i j} \mathbf{R}_{i j} c_{j s}^{\dagger} c_{i s}, \quad \hat \tau=e^2\sum_{i, j, s} h_{i j} \mathbf{R}_{i j} \otimes \mathbf{R}_{i j} c_{j s}^{\dagger} c_{i s}.
\end{equation}
Finally, we can expand the current operator in the interaction picture into a series of ${\bf A}(t)$,
\begin{equation}
	\hat{\boldsymbol{j}}_I(t)=-\frac{\delta \hat{H}_I}{\delta \boldsymbol{A}}(t)=e^{iH_0t}\left\{
	\hat{\boldsymbol{j}}_0-\hat{\tau} \mathbf{A}(t) + \dots\right\}e^{-iH_0t}.
\end{equation}

Now, we start from the zeroth order current in Eq.~\eqref{eq:current_order}, where $\hat{\rho}^{(0)}(t) = \hat{\rho}(0)$ commutes with $e^{-iH_0t}$:
\begin{equation}
	\boldsymbol{j}^{(0)}(t) = \tr\left(\hat{\boldsymbol{j}}_I(t)\hat{\rho}^{(0)}(t)\right) = \langle\boldsymbol{j}_0\rangle-
	\langle\hat{\tau}\rangle\boldsymbol{A}(t)+\mathcal{O}(\mathbf{A}^2(t) ). 
\end{equation}
Here, we have used the notation $\langle \dots\rangle\equiv \tr(\dots\rho(0))$. The first term vanishes when the system has no current before the perturbation. The second term vanishes when the vector potential is zero at time $t$.
Next, for the first order current,
\begin{equation}
	\begin{aligned}
		\boldsymbol{j}^{(1)}(t) &= \tr\left(\hat{\boldsymbol{j}}_I(t)\hat{\rho}^{(1)}(t)\right) = \tr\left(\hat{\boldsymbol{j}}_I(t)(-i)\int_0^t dt_1[\hat{H}'_I(t_1),\hat{\rho}(0)]\right)\\
		&=-i\int_0^t dt_1\tr\left(\hat{\boldsymbol{j}}_{0I}(t)\left[\hat{\boldsymbol{j}}_{0I}(t_1)\boldsymbol{A}(t_1),\rho(0)\right]\right)+\mathcal{O}\left(\boldsymbol{A}(t)\boldsymbol{A}(t_1)+\boldsymbol{A}^2(t_1)\right)\\
		&=-i\int_0^t dt_1\Big\langle\Big[\hat{\boldsymbol{j}}_{0I}(t),\hat{\boldsymbol{j}}_{0I}(t_1)\Big]\Big\rangle\boldsymbol{A}(t_1)+\mathcal{O}\left(\boldsymbol{A}(t)\boldsymbol{A}(t_1)+\boldsymbol{A}^2(t_1)\right).
	\end{aligned}
\end{equation}
Within linear response theory, one defines the two-time optical conductivity 
\begin{equation}
	\sigma\left(t, t_1\right)\equiv\left\langle\tau\right\rangle_t-\int_0^t \chi\left(t, t_1\right) \mathrm{d} t_1 \quad (t>t_1>0)
\end{equation}
with susceptibility
\begin{equation}
	\chi(t,t_1)=i\Big\langle \Big[\hat{\boldsymbol{j}}_{0I}(t),\hat{\boldsymbol{j}}_{0I}(t_1)\Big]\Big\rangle.
\end{equation} 
Taking into account $ \boldsymbol{A}(t_1)=-\int_0^{t_1} \boldsymbol{E}(\bar{t})\mathrm{d}\bar{t}$,  the linear current $\boldsymbol{j}(t)=\int_{0}^{t}\mathrm{d} t_1\sigma(t,t_1)\boldsymbol{E}(t_1)$ is thus given by the zeroth and first order contribution above.

Similarly, for the second order current, we find
\begin{equation}
	\begin{aligned}
		\boldsymbol{j}^{(2)}(t)  &=\tr\left(j_I(t)\rho^{(2)}(t)\right)= \tr\left(j_I(t)(-i)^2\int_0^{t_1} dt_{2}\int_0^t dt_1[H'_I(t_{2}),[H'_I(t_1),\rho(0)]]\right)\\
		&=(-i)^2\int_0^{t_1} dt_2\int_0^t dt_1\tr\left(\hat{\boldsymbol{j}}_{0I}(t)\left[\hat{\boldsymbol{j}}_{0I}(t_{2})\boldsymbol{A}(t_2),\left[\hat{\boldsymbol{j}}_{0I}(t_1)\boldsymbol{A}(t_1),\rho(0)\right]\right]\right)+\mathcal{O}(\boldsymbol{A}^3 )\\
		&=(-i)^2\int_0^{t_1} dt_2\int_0^t dt_1\Big\langle\left[\hat{\boldsymbol{j}}_{0I}(t),\left[\hat{\boldsymbol{j}}_{0I}(t_{2})\boldsymbol{A}(t_2),\hat{\boldsymbol{j}}_{0I}(t_1)\boldsymbol{A}(t_1)\right]\right]\Big\rangle+\mathcal{O}(\boldsymbol{A}^3 ),
	\end{aligned}
\end{equation}
and for the third order current 
\begin{equation}
	\begin{aligned}
		\boldsymbol{j}^{(3)}(t) &= \tr\left(\hat{\boldsymbol{j}}_I(t)\rho^{(3)}(t)\right) = \tr\left(\hat{\boldsymbol{j}}_I(t)(-i)^3\int_0^t dt_1\int_0^{t_1} dt_2\int_0^{t_2} dt_3[\hat{H}'_I(t_3),[\hat{H}'_I(t_{2}),[\hat{H}'_I(t_1),\hat{\rho}(0)]]]\right)\\
		& =(-i)^3\int_0^t dt_1\int_0^{t_1} dt_2\int_0^{t_2} dt_3\tr
		\Big\langle\left[\hat{\boldsymbol{j}}_{0I}(t),\left[\hat{\boldsymbol{j}}_{0I}(t_{3})\boldsymbol{A}(t_3),\left[\hat{\boldsymbol{j}}_{0I}(t_{2})\boldsymbol{A}(t_2),\hat{\boldsymbol{j}}_{0I}(t_1)\boldsymbol{A}(t_1)\right]\right]\right]\Big\rangle+\mathcal{O}(\boldsymbol{A}^4 ).
	\end{aligned}
\end{equation}

\section{Keldysh contour formalism (Liouville paths, double-sided Feymann diagram)}\label{sec:keldysh}

The Keldysh formalism provides a convenient framework for representing the nonequilibrium evolution of the density matrix.
In the Schr\"{o}dinger  picture, the observable corresponding to the operator $\hat{O}$ is

$$O(t) = \tr\left(\hat{\rho}(0) \hat{O}(t) \right)=\tr\left[\mathscr{U}\left(t, t_0\right)\hat{\rho}(0) \mathscr{U}\left(t, t_0\right)^{\dagger} \hat{O} \right]
$$
with 
$
\mathscr{U}\left(t, t_0\right) = \mathcal{T}_t \exp [-i \int_{t_0}^t d \bar{t} \hat{H}(\bar{t})].
$
In the Keldysh path integral formalism, one can separate the time dependent part $H'(t)$ from the time independent part $H_0(t)$, 
\begin{equation}
	\begin{aligned}
		O(t)&=\frac{1}{Z} \tr\left[\mathcal{T}_{\mathcal{C}} e^{-i \int_{\mathcal{C}} d \bar{t} \hat{H}_0}  e^{-i \int_{\mathcal{C}} d \bar{t} \hat{H}'(\bar{t})} \hat{O}\left(t_{+}\right)\right]\\
		&=\frac{1}{Z} \tr\left[\mathcal{T}_{\mathcal{C}} e^{-i \int_{\mathcal{C}} d \bar{t} \hat{H}_0}  e^{-i \int_{\mathcal{C}_-} d t_- \hat{H}'(t_-)} \hat{O}\left(t_{+}\right) e^{-i \int_{\mathcal{C}_+} d t_+ \hat{H}'(t_+)}\right]\\
		&=\frac{1}{Z} \tr\left[\mathcal{T}_{\mathcal{C}} e^{-i \int_{\mathcal{C}} d \bar{t} \hat{H}_0} \left(\sum_{n=0}^\infty (-i)^n  \int_{\mathcal{C}^-}d t^-_n\dots d t^-_1\hat{H}'(t^-_n) \dots\hat{H}'(t^-_1) \right)\hat{O}\left(t^{+}\right) \right.\\
		&\quad\quad\quad\quad\quad\quad\quad\quad\quad\quad\left.\left(\sum_{n=0}^\infty (-i)^n \int_{\mathcal{C}^+}d t^+_n\dots d t^+_1\hat{H}'(t^+_n) \dots\hat{H}'(t^+_1) \right) \right].
	\end{aligned}
\end{equation}
Collecting the orders of $H'$, one obtains the zero-th, first and second order electric current response,
\begin{equation}
	\boldsymbol{j}^{(0)}(t)=\frac{1}{Z} \tr\left[T_{\mathcal{C}} e^{-i \int_{\mathcal{C}} d \bar{t} \hat{H}_0(\bar{t})} \hat{\boldsymbol{j}}\left(t^{+}\right)\right],
\end{equation}
\begin{equation}
	\boldsymbol{j}^{(1)}(t)=\frac{(-i)}{Z} \tr\left[T_{\mathcal{C}} e^{-i \int_{\mathcal{C}}  d \bar{t} \hat{H}_0(\bar{t})}\left( \hat{\boldsymbol{j}}\left(t^{+}\right)  \int_{\mathcal{C}^+}d t^+_1\hat{H}'(t^+_1)+
	\int_{\mathcal{C}^-}d t^-_1\hat{H}'(t^-_1) \hat{\boldsymbol{j}}\left(t^{+}\right) ,
	\right)\right]
\end{equation}
\begin{equation}
	\begin{aligned}
		\boldsymbol{j}^{(2)}(t)=\frac{(-i)^2}{Z} &\tr\left[T_{\mathcal{C}} e^{-i \int_{\mathcal{C}}  d \bar{t} \hat{H}_0(\bar{t})}\left( \hat{\boldsymbol{j}}\left(t^{+}\right)  \iint_{\mathcal{C}^+}dt^+_2t^+_1\hat{H}'(t^+_2)\hat{H}'(t^+_1)+
		\int_{\mathcal{C}^+} d t^+_2\int_{\mathcal{C}^-}d  t^-_1\hat{H}'(t^-_2) \hat{\boldsymbol{j}}\left(t^{+}\right) \hat{H}'(t^-_1)\right.\right.\\
		&\left.\left.
		+\int_{\mathcal{C}^+} d t^+_1\int_{\mathcal{C}^-}d t^-_2\hat{H}'(t^-_1) \hat{\boldsymbol{j}}\left(t^{+}\right) \hat{H}'(t^-_2)+
		\iint_{\mathcal{C}^-}d t^-_2d t^-_1\hat{H}'(t^-_2)\hat{H}'(t^-_1) \hat{\boldsymbol{j}}\left(t^{+}\right) 
		\right)\right].
	\end{aligned}
\end{equation}
We are interested in the third order current $\boldsymbol{j}^{(3)}(t)$. Expanding $\hat{H}'$ in powers of $\boldsymbol{A}$ in the weak field limit gives $\hat{H}'(t) = \hat{\boldsymbol{j}}(t)\boldsymbol{A}(t) +\mathcal{O}({\bf A}^2)$. The third order current thus includes eight terms, which we can write as
\begin{equation}
	\begin{aligned}
		\boldsymbol{j}^{(3)}(t)=\frac{(-i)^3}{Z} &\tr\left[T_{\mathcal{C}} e^{-i \int_{\mathcal{C}}  d \bar{t} \hat{H}_0(\bar{t})}   \iiint_{\mathcal{C}_{+/-}} dt_3dt_2dt_1 \boldsymbol{A}(t_3)\boldsymbol{A}(t_2)\boldsymbol{A}(t_1)R^{(3)}(t,t_3,t_2,t_1)\right]+\mathcal{O}(A^4), 
	\end{aligned}
\end{equation}
with the four point response ``generator" ($t>t_3>t_2>t_1$)
\begin{equation}
	\begin{aligned} 
		R^{(3)}(t,t_3,t_2,t_1) = 
		&
		\hat{\boldsymbol{j}}_{1-}\hat{\boldsymbol{j}}_{3-}\hat{\boldsymbol{j}}_{t}\hat{\boldsymbol{j}}_{2+} +		\hat{\boldsymbol{j}}_{2-}\hat{\boldsymbol{j}}_{t}\hat{\boldsymbol{j}}_{3^+}\hat{\boldsymbol{j}}_{1+} \\
		&		\hat{\boldsymbol{j}}_{1-}\hat{\boldsymbol{j}}_{2-}\hat{\boldsymbol{j}}_{t}\hat{\boldsymbol{j}}_{3+} +		\hat{\boldsymbol{j}}_{3^-}\hat{\boldsymbol{j}}_{t}\hat{\boldsymbol{j}}_{2+}\hat{\boldsymbol{j}}_{1^+} \\
		&		\hat{\boldsymbol{j}}_{2-}\hat{\boldsymbol{j}}_{3-}\hat{\boldsymbol{j}}_{t}\hat{\boldsymbol{j}}_{1+} +		\hat{\boldsymbol{j}}_{1-}\hat{\boldsymbol{j}}_{t}\hat{\boldsymbol{j}}_{3+}\hat{\boldsymbol{j}}_{2+} \\
		&		\hat{\boldsymbol{j}}_t\hat{\boldsymbol{j}}_{3+}\hat{\boldsymbol{j}}_{2+}\hat{\boldsymbol{j}}_{1+} +		\hat{\boldsymbol{j}}_{1-}\hat{\boldsymbol{j}}_{2-}\hat{\boldsymbol{j}}_{3-}\hat{\boldsymbol{j}}_t .\\
	\end{aligned}
\end{equation}
$\hat{\boldsymbol{j}}_{i\pm}$ is a short hand notation for $\hat{\boldsymbol{j}}(t_i)$, $t_i\in \mathcal{C}_\pm$. The Keldysh index of $t$  is dropped due to the fact that $\hat{\boldsymbol{j}}(t^+)=\hat{\boldsymbol{j}}(t^-)$ at the end of the real time branches of the Keldysh contour. The time integration on the $\mathcal{C}_-$ branch brings a negative sign compared to a normal time integral. 

For simplicity, we consider the semi-impulsive limit, i.e., assume an external field $\boldsymbol{A}(t) \propto \delta(t) \cos(\omega t -\boldsymbol{k}\cdot \boldsymbol{r})= \delta(t) \left(e^{-i\omega t +i\boldsymbol{k}\cdot \boldsymbol{r}}+e^{i\omega t -i\boldsymbol{k}\cdot \boldsymbol{r}}\right) \hat{\boldsymbol{r}}$ which is the sum of an excitation and deexcitation. If a given pulse would create additional excitations and deexcitations along the contour, it would complicate the analysis.

In the non-collinear setup, we can selectively detect the signal in a given direction. For example, in the ``box" geometry of four-wave-mixing, after three non-collinear pump pulses, we can put the fourth pump (local oscillator) along the direction $\vec{k}_\text{sig} = ( -\vec{k}_A+\vec{k}_B+\vec{k}_C)$. By ordering the time sequences of the laser pulses $A$, $B$, $C$, we can select the excitation or deexcitation term of the light-matter interaction.

\begin{figure}[h]
	\centering
	\includegraphics[width=0.99\linewidth]{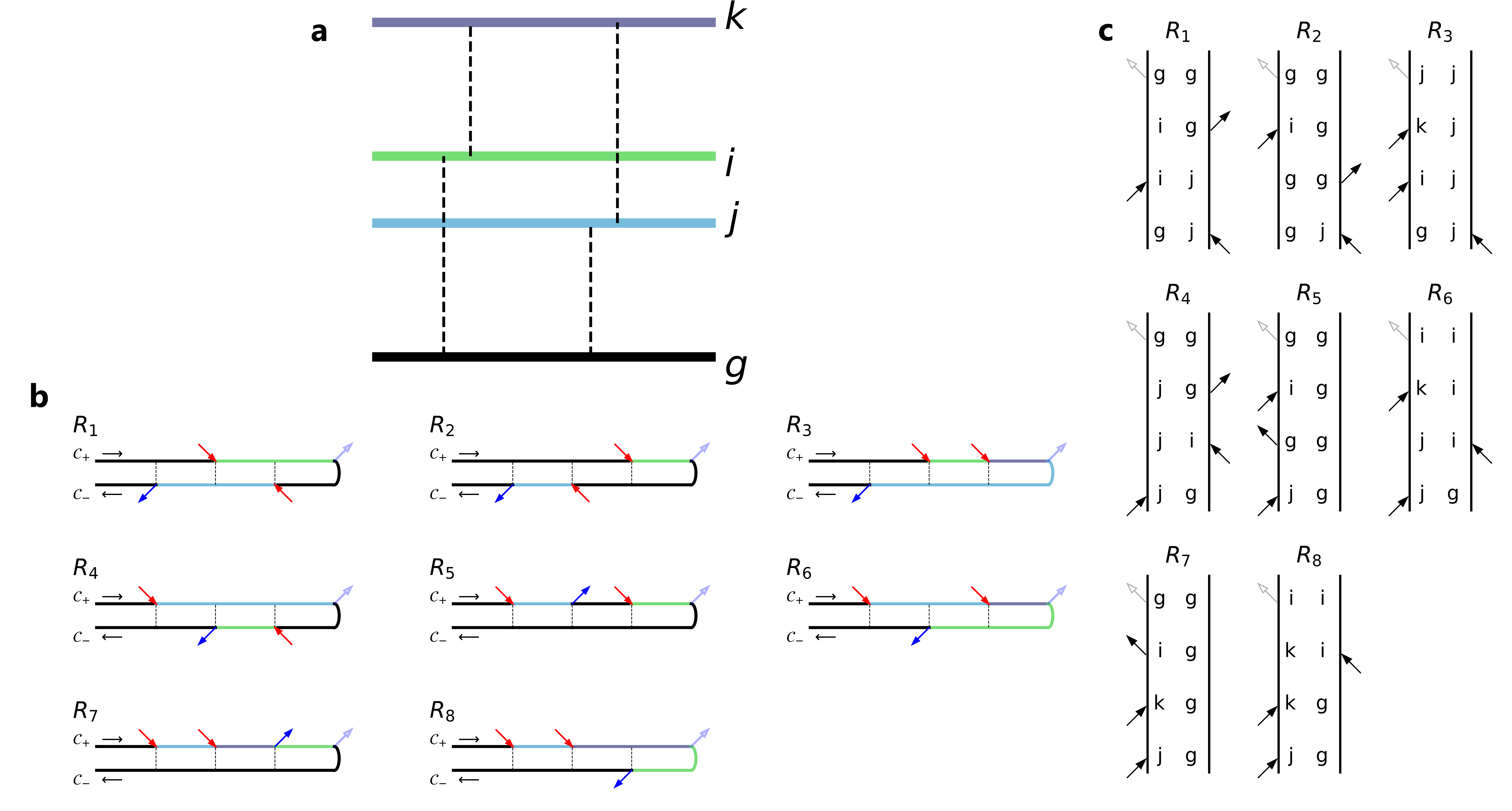}
	\caption{{\bf a}, Energy levels of a generic quantum system. {\bf b}, Keldysh contours for the R (top), NR (middle) and 2Q (bottom) signals $R_1$-$R_8$. {\bf c}, Double-sided Feymann diagrams for the same signals as in {\bf b}.}
	\label{fig:si}
\end{figure}

Following the notations in~Ref.~\onlinecite{hamm2011}, if $k_A$ comes first, the $t_1$ pulse deexcites, while $t_2$, $t_3$ excites along the contour. Three of the pathways survive, the rephasing diagrams $R_1$, $R_2$, and $R_3$ (top panels of Figure~\ref{fig:si}{\bf b}). If $k_B$ comes first, the $t_2$ pulse excites while $t_2$, $t_3$ deexcites, as illustrated by the 
nonrephasing diagrams $R_4$, $R_5$, $R_6$ in the middle panels of Figure~\ref{fig:si}{\bf b}. If $k_C$ comes first, $t_3$ excites and $t_1$, $t_2$ deexcites, as in the two 2Q pathways $R_7$, $R_8$ in the bottom panels of Figure~\ref{fig:si}{\bf b}. 

We note that in the presence of inhomogeneous broadening of the excitations, the R signals ($R_1$ ,$R_2$, $R_3$) rephase when $t=\tau$, while the NR signals ($R_4$ ,$R_5$, $R_6$) keep dephasing. Consider an excitation energy centered at $\omega_0$ with Gaussian broadening of width $\Delta\omega$. Neglecting the decoherence during the time evolution, the R and NR signals are
$	S^{\text{R/NR}}(\tau,t) $. 
The R signal echos when $t=\tau$, since the factor $e^{-\left(\tau- t\right)^2 \Delta \omega^2/2} =1$, while the NR signal with $e^{-\left(\tau+ t\right)^2 \Delta \omega^2/2}$ keeps decreasing.

For comparison, we also plot the traditional double-sided Feymann  diagrams in Figure~\ref{fig:si}{\bf c}. In Table~\ref{table2}, the signal locations of each pathway in the $(\omega_\tau,\omega_t)$ domains are listed, together with the frequencies of the intensity oscillations during $T$.

\begin{table}[!h]
\begin{center}
	\begin{tabular}{ | >{\centering}p{5mm}  | >{\centering}p{8mm}| >{\centering}p{2cm}| c|} 
		\hline
		&signal &peak location  & oscillation    \\ 
		&index &($\omega_\tau$,$\omega_t$) & frequency $\Omega$  \\ \hline\hline
		
		\multirow{3}*{R}
		&$R_1$ &$(-\omega_j,\omega_i)$ & $\omega_{ij}$   \\ \cline{2-4}
		&$R_2$ &$(-\omega_j,\omega_i)$ & $\omega_{ij}$    \\ \cline{2-4}
		&$R_3$ &$(-\omega_j,\omega_{jk})$ & $\omega_{ij}$    \\ \hline\hline
		\multirow{3}*{NR}
		&$R_4$ &$(\omega_j,\omega_j)$ & $\omega_{ij}$   \\ \cline{2-4}
		&$R_5$ &$(\omega_j,\omega_i)$ & 0 \\ \cline{2-4}
		&$R_6$ &$(\omega_j,\omega_{ik})$ & $\omega_{ij}$   \\ \hline\hline
		\multirow{2}*{2Q}
		&$R_7$ &$(\omega_j,\omega_i)$ & $\omega_k$   \\ 
		\cline{2-4}
		&$R_8$ &$(\omega_j,\omega_{ik})$ & $\omega_k$  \\ \hline
		
	\end{tabular}
	
	\caption{Locations of the R (top), NR (middle) and 2Q (bottom) signals in the $(\omega_\tau,\omega_t)$ plane and oscillation frequencies of the signal intensities during the waiting time $T$.}\label{table2}
	
\end{center}
\end{table}

\end{document}